\documentclass{mn2e}

\usepackage{amsfonts}
\usepackage{amsmath}
\usepackage{graphicx}

\title[Nonlinear dark matter pdf]
      {Perturbation theory and excursion set estimates of the 
       probability distribution function of dark matter, and 
       a method for reconstructing the initial distribution function}

\author[T. Y. Lam \& R. K. Sheth]
{Tsz Yan Lam\thanks{E-mail:  tylam@sas.upenn.edu, shethrk@physics.upenn.edu}
 \& Ravi K. Sheth\footnotemark[1]\\
 Department of Physics \& Astronomy, University of Pennsylvania, 
 209 S. 33rd Street, Philadelphia, PA 19104, USA}

\newcommand{\bm}[1]{{\mbox{\boldmath $#1$}}}

\begin{document}
\pagerange{\pageref{firstpage}--\pageref{lastpage}}

\maketitle

\label{firstpage}

\begin{abstract}
Nonlinear evolution is sometimes modeled by assuming there is a 
deterministic mapping from initial to final values of the 
locally smoothed overdensity.  
However, if an underdense region is embedded in a denser one, then 
it is possible that its evolution is determined by its surroundings, 
so the mapping between initial and final overdensities is not as 
`local' as one might have assumed.  If this source of nonlocality 
is not accounted for, then it appears as stochasticity in the mapping 
between initial and final densities.  
Perturbation theory methods ignore this `cloud-in-cloud' effect, 
whereas methods based on the excursion set approach do account 
for it; as a result, one may expect the two approaches to provide 
different estimates of the shape of the nonlinear counts in cells 
distribution.  We show that, on scales where the rms fluctuation is 
small, this source of nonlocality has only a small effect, so the 
predictions of the two approaches differ only on the small scales on 
which perturbation theory is no longer expected to be valid anyway.  
We illustrate our results by comparing the predictions of these 
approaches when the initial-final mapping is given by the spherical 
collapse model.  Both are in reasonably good agreement with measurements 
in numerical simulations on scales where the rms fluctuation is 
of order unity or smaller.  

If the deterministic mapping from initial conditions to final density 
depends on quantities other than the initial density, then this will 
also manifest as stochasticity in the mapping from initial density to 
final.  For example, the Zeldovich approximation and the 
Ellipsoidal Collapse model both assume that the initial shear field 
plays an important role in determining the evolution.  
We compare the predictions of these approximations with simulations, 
both before and after accounting for the `cloud-in-cloud' effect.  
Our analysis accounts approximately for the fact that the shape of a 
cell at the present time is different from its initial shape; ignoring 
this makes a noticable difference on scales where the rms fluctuation 
in a cell is of order unity or larger.  

On scales where the rms fluctuation is 2 or less, methods based on 
the spherical model are sufficiently accurate to permit a rather 
accurate reconstruction of the shape of the initial distribution 
from the nonlinear one.  This can be used as the basis for a method 
for constraining the statistical properties of the initial fluctuation 
field from the present day field, under the hypothesis that the 
evolution was purely gravitational.  We illustrate by showing how 
the highly non-Gaussian nonlinear density field in a numerical 
simulation can be transformed to provide an accurate estimate of the 
initial Gaussian distribution from which it evolved. 
\end{abstract}


\section{Introduction}
The present work is primarily concerned with the probability 
distribution function (hereafter pdf; some authors prefer to call 
this the probability density function), which describes the 
probability that a randomly placed cell of specified shape and 
volume contains a certain amount of mass.  In the best studied 
cosmological models, the pdf has a Gaussian form initially, but 
becomes increasingly positively skewed at later times.  
There are two methods for estimating the evolution of the dark 
matter pdf.  One is based on perturbation theory 
\cite{b94,ps97,gc99,sg01,ptreview} 
and the other is based on excursion set methods \cite{rks98}.  
The perturbation theory based calculation is not expected to be 
accurate on scales where the variance is of order unity or larger; 
in hierarchical models, this means that perturbation theory is 
not expected to be accurate on small scales.  
Hyperextended Perturbation Theory (HEPT) \cite{hept} is expected to be 
valid in the larger variance (small cell) regime where standard 
perturbation theory breaks down; it provides explicit expressions 
for the moments of the pdf rather than a closed form expression.  
For the special case of clustering from white-noise Gaussian initial 
conditions, the excursion set method predicts exactly the same pdf 
as does HEPT, despite the fact that 
the methods used by the two approaches are very different.  
Motivated by this coincidence, the purpose of the present note is 
twofold:  first, to show that, for more general initial conditions, 
the excursion set approach actually makes rather similar predictions 
to those of perturbation theory on scales where the variance is small; 
the second is to understand why.  

In the analysis which follows, we distinguish between two steps 
in the calculation of the pdf.  The first is the approximation for 
nonlinear evolution which we will call the dynamics.  The second is 
how this approximation is used to translate from the initial pdf to 
an evolved one, which we call the statistics.  In the first half of 
this paper, we study approximations for the dynamics which are based 
on the assumption of spherical symmetry.  
In this case, the perturbation theory method provides what is, in 
effect, a monotonic, deterministic mapping between the initial and 
final overdensities.  Because the final overdensity at a specified 
position in space is determined solely by the initial value at that 
position, this is sometimes also called a `local' mapping, since 
values of the initial fluctuation field at other positions are assumed 
to not affect the mapping.  

The excursion set approach accounts for the fact that the evolution 
of a given region may actually be determined by less local surroundings.  
For example, consider the evolution of an underdense region which 
is surrounded by a dense shell.  If the shell is sufficiently dense, 
then it will eventually collapse, crushing the smaller region within 
it.  The local approximation would have predicted expansion rather 
than collapse for the smaller underdense region.  
Sheth \& van de Weygaert (2004) call this the void-in-cloud problem, 
although it is clear that this is an extreme example of a more 
general `cloud-in-cloud' problem.  Clearly, in such cases, the mapping 
between initial and final overdensities is not as `local' as 
perturbation theory assumes, and accounting for this `cloud-in-cloud' 
problem is likely to be more important for small `clouds'.  
If not accounted for, this effect will manifest both as stochasticity 
(since the same initial overdensity may map to many different final 
densities depending on the surroundings) and, perhaps, as a bias.  
The excursion set approach provides an algorithm which accounts for 
this source of non-locality; it assumes that, once the correct large 
scale has been chosen, the mapping is deterministic.

However, there is another source of non-locality which spherical 
evolution models ignore.  This source plays a crucial role in models 
which account for the influence of the external shear field on the 
evolution.  The simplest of these more sophisticated models for the 
dynamics is the Zeldovich approximation \cite{zel70}.  Here, the 
nonlinear density is determined not just by the initial overdensity, 
but by two other quantities which are related to the surrounding 
shear field.  These quantities also play an important role in 
extensions of the Zeldovich approximation \cite{mkc01} as well 
as in ellipsoidal collapse models \cite{ws79,bm96}.
In all these models, the nonlinear density is a deterministic function 
of three quantities associated with the initial fluctuation field.  
In the context of perturbation theory models for the pdf, the mapping 
from initial density to final density will appear to be stochastic 
if the influence of the two other variables is not accounted for.  
In the excursion set approach, this stochasticity is in addition to 
that which derives from the cloud-in-cloud problem which is now 
associated with all three variables.  

To explore the consequences of this additional source of stochasticity 
we show the result of inserting the Ellipsoidal collapse model into 
the perturbation theory and excursion set calculations.  We do this 
in two steps:  by showing the predictions when expanding the dynamics 
to first, and then to second order.  To first order, the Ellipsoidal 
collapse model reduces to the Zeldovich approximation; hence, our   
perturbation theory discussion of the first order approximation 
extends previous work on the Zeldovich approximation \cite{kbg94,hks00}.  
However, our analysis also accounts for another subtlety associated 
with non-spherical collapse models---that the final shape of a patch 
is different from its initial shape \cite{br91,pk93,brlc02}.  
We then present the results of the simplest excursion set treatment of 
this problem which accounts for the evolution in the volumes but not 
the shapes of cells.  

The perturbation theory and excursion set predictions associated 
with spherical dynamics are presented in Section~\ref{models}.  
This section also contains a discussion of the Lognormal model.  
It includes a comparison of these predictions with measurements 
in simulations.  
The perturbation and excursion set treatments of the Zeldovich and 
ellipsoidal collapse models are in Section~\ref{stochastic}. 
A final section summarizes our findings, and includes a discussion 
of the fact that local deterministic mapping models of nonlinear 
evolution motivate a simple method for reconstructing the shape of 
the initial pdf from that at late times.  We illustrate the method 
using the spherical evolution model; we also show that, for the 
present purposes, the spherical model is a good enough approximation 
to the second order ellipsoidal collapse dynamics.

\section{Deterministic mappings from initial to final density}\label{models}
This section describes the perturbation theory and excursion set 
models of the pdf when the mapping from linear to nonlinear density 
is deterministic.  In both, the variance of the initial density 
fluctuation field when smoothed on scale $R_M$ plays an important 
role.  It is denoted by 
\begin{equation}
 \sigma_{\rm L}^2(M) \equiv
 \int \frac{{\rm d}k}{k}\,\frac{k^3\,P_{\rm L}(k)}{2\pi^2}\, |W(kR_M)|^2
 \label{sigmaL}
\end{equation}
where $P_{\rm L}(k)$ is the power spectrum of the initial field, 
$W(x)$ is the Fourier transform of the smoothing window, and 
$R_M \equiv (3M/4\pi\bar\rho)^{1/3}$.
We will also use the quantity 
\begin{equation}
\gamma \equiv -3\,\frac{{\rm d}\ln \sigma^2_L}{{\rm d}\ln M}.
\label{eqn:gamma}
\end{equation}
For $P(k)\propto k^n$, $\gamma = (n+3)$.

\subsection{Perturbation theory-based methods}\label{subsection:PT}
These methods make three important assumptions.  
First, an initial overdensity $\delta_{\rm L}$ can be mapped to an 
evolved density $\rho$.  
Because $\rho$ at any given position in space is determined from 
$\delta_{\rm L}$ at the same position (in Lagrangian coordinate), 
the mapping is said to be 
`local'.  Because $\rho$ depends on $\delta_{\rm L}$ and nothing 
else, the mapping is said to be `deterministic'.  
Notice that these `local' `deterministic' assumptions make no 
mention of the scale on which they apply.  In what follows, we 
will write the evolved density in a cell of volume $V$, in units 
of the background density, as 
\begin{equation}
\rho \equiv \frac{M}{\rho_bV} \equiv 1 + \delta;
\end{equation}
we hope that this slight abuse of notation will not cause confusion.
Since the evolved density is clearly smoothed on scale $V$, 
the question is:  on what smoothing scale should the initial 
overdensity $\delta_{\rm L}$ associated with $\rho$ be defined?  
This is where the third key assumption is made:  the appropriate 
scale is that which initially contains mass $M$.  As a result of 
this assumption, the cumulative distributions of the evolved 
(Eulerian) pdf at fixed $V$ and the initial (Lagrangian) pdf at 
fixed mass scale $M$ are related as follows (see Section 5.4.3 
in Bernardeau et al. 2002):
\begin{equation}
 \int_{M}^{\infty} {\rm d}M\,p(M|V)\,{M\over \bar M} = 
 \int_{\delta_{\rm L}(M|V)/\sigma_{\rm L}(M)}^\infty
   {\rm d}x\,\frac{\exp(-x^2/2)}{\sqrt{2\pi}},
 \label{eqn:cumulative}
\end{equation}
where $\bar M\equiv\int_0^{\infty} {\rm d}M\,p(M|V)\,M$,
$\sigma_{\rm L}$ is given by equation~(\ref{sigmaL}), and 
the right hand side assumes that the initial distribution was 
Gaussian.  
Differentiating with respect to $M$ yields 
\begin{equation}
 {M\over \bar M}\,p(M|V) = 
 \frac{\exp[-(\delta_{\rm L}/\sigma_{\rm L})^2/2]}{\sqrt{2\pi}}
 \frac{{\rm d}\,(\delta_{\rm L}/\sigma_{\rm L})}{{\rm d}M}.
\end{equation}
Since $M/\bar M\equiv \rho$, 
the expression above implies that 
\begin{equation}
 \rho^2\,p(\rho|V) = 
 \exp\left[-\frac{\delta_{\rm L}^2}{2\sigma_{\rm L}^2}\right]
 \sqrt{\frac{\delta_{\rm L}^2}{2\pi\sigma_{\rm L}^2}}\,
 \frac{{\rm d}\ln(\delta_{\rm L}/\sigma_{\rm L})}{{\rm d}\ln \rho}.
 \label{eqn:PTpdf}
\end{equation}
The requirement that the left hand side of equation~(\ref{eqn:cumulative}) 
decreases monotonically with increasing $M$ (since we want $p(M|V)\ge 0$ 
for all $M$) means that $\delta_{\rm L}/\sigma_{\rm L}$ must increase 
monotonically with increasing $M$.  As a result, the range of allowed 
$\delta_{\rm L}(M|V)$ relations is constrained by the relation 
$\sigma_{\rm L}(M)$, i.e., by the initial power spectrum.

\subsection{Excursion set method}\label{subsection:randomwalk}
The excursion set model \cite{rks98} exploits the fact that 
$\sigma_{\rm L}^2(M)$ is a monotonic function of $M$, so the local 
collapse model $\delta_{\rm L}(M|V)$ defines a curve in the space 
of $\delta_{\rm L}$ versus $\sigma_{\rm L}^2$.  In what follows, 
we will set $S(M) = \sigma_{\rm L}^2(M)$.  Then, the excursion 
set model for the distribution of $M$ in cells of size $V$ is 
\begin{equation}
 \rho^2 p(\rho|V) = Sf(S|V) \frac{d \ln S}{d \ln \rho} 
 \label{eqn:RWpdf}
\end{equation}
where $f(S|V)\,{\rm d}S$ denotes the probability that a random walk 
with uncorrelated steps first crosses a barrier of height $B(S|V)$ on 
scale $S$ (where $S$ denotes the variance in walk heights).  
The collapse dynamics is included in this solution of the statistical 
problem by setting $B(S|V) = \delta_{\rm L}(\sigma_{\rm L}^2)$.  
It is by relating the counts in cells distribution to the first 
crossing distribution that the excursion set model accounts for 
the `cloud-in-cloud' problem discussed in the Introduction.  Note 
that this method allows a larger range of $\delta_{\rm L}(M|V)$ 
relations than does perturbation theory.  

We have used two methods for computing $f(S|V)\,{\rm d}S$: 
one is a Monte-Carlo approach, where we simulate a large ensemble of 
random walks and count the distribution of first crossings, and the 
other is the analytical approximation of Sheth \& Tormen (2002).  
This approximation sets 
\begin{equation}
f(S|V)dS = |T(S|V)|\,\exp\left[-\frac{B(S|V)^2}{2S}\right]
             \frac{dS/S}{\sqrt{2\pi S}},
\end{equation}
where $T(S|V)$ denotes the first few terms of the Taylor expansion 
of $B$ about $S$.  
For the barrier shapes of interest in this paper, including only 
the first two terms is sufficient, so 
\begin{equation}
 T(S|V) \approx \,
  B(S|V)\left[1 - \frac{\partial \ln B(S|V)}{\partial \ln S}\right],
\end{equation}
where the derivative is evaluated at $S$.  Thus, 
\begin{eqnarray}
 \rho^2 p(\rho|V) &=& \exp\left[-\frac{B(S|V)^2}{2S}\right]
             \sqrt{\frac{B(S|V)^2}{2\pi S}}
             \frac{d \ln S}{d \ln \rho} \nonumber\\
       && \qquad \times\qquad
             \left| 1 - \frac{\partial \ln B(S|V)}{\partial \ln S}\right|
             \nonumber\\
    &=& \exp\left[-\frac{\delta_{\rm L}^2}{2\sigma_{\rm L}^2}\right]
         \sqrt{\frac{\delta_{\rm L}^2}{2\pi \sigma_{\rm L}^2}}
         \left|\frac{d\ln (\delta_{\rm L}/\sigma_{\rm L}^2)}{d\ln \rho}\right|,
\label{eqn:RWapprox}
\end{eqnarray}
where the final expression uses the fact that 
$B^2/S \equiv (\delta_{\rm L}/\sigma_{\rm L})^2$.  
Comparison with perturbation theory (equation~\ref{eqn:PTpdf}) shows 
that the only difference is in the Jacobian like term, which differs 
by a factor of $\sigma_{\rm L}$.

\subsection{Normalization}\label{norm}
Direct implementation of the two methods described above produces 
distributions which are guaranteed to have the correct mean value, 
$\int d\rho\,p(\rho|V)\,\rho = 1$, but, in general, 
$\int d\rho\,p(\rho|V) \ne 1$.  
(Strictly speaking, the analytic approximation to the random 
walk does not integrate to unity; the Monte-Carlo solution, of course, 
does; nevertheless the analytical formula gives a very good approximation.) 
To ensure that this integral also equals unity, one must define 
\begin{equation}
 N \equiv \int_0^{\infty} d\rho\,p(\rho), \ 
 \rho' \equiv N\rho, \quad {\rm and}\
 \rho'^2 p'(\rho') = \rho^2 p(\rho).
\end{equation}
The quantities $\rho'$ and $p'(\rho')$ now satisfy both normalisation 
requirements:
 $\int d\rho'\,p'(\rho') = 1$ and $\int d\rho'\, p'(\rho')\,\rho'=1$. 
Note that although this procedure is standard \cite{kbg94,ps97,ecpdf}, 
this procedure has no physical motivation, other than to insure 
correct normalization.

The last of the expressions above shows that plots of $\rho^2\,p(\rho)$ 
versus log$(\rho)$ can be turned into properly normalized plots 
simply by shifting along the log$(\rho)$ direction.  This is useful 
because, as equations~(\ref{eqn:PTpdf}) and~(\ref{eqn:RWpdf}) show, 
it is $\rho^2\,p(\rho)$ which is the fundamental prediction of 
both models for the statistics.  We will use this in what follows.

\subsection{Example:  The Lognormal distribution}
The standard form of the Lognormal distribution is 
\begin{equation}
 \rho^2\, p(\rho)
 = {\exp[-(\ln\rho + \mu)^2/2\sigma^2]\over \sqrt{2\pi}} 
   \,\frac{\rho}{\sigma}
 \label{eqn:pln}
\end{equation}
where $\mu=\sigma^2/2$ and 
$\sigma^2 = \ln \langle\rho^2\rangle$. 
This can be scaled to a Gaussian distribution in the variable 
$\delta_{\rm L}$ with mean zero and variance $\sigma_{\rm L}$ by 
setting $\rho \equiv 1+\delta \propto \exp(\delta_{\rm L})$ so 
$\ln(1+\delta) \propto \delta_{\rm L}$ where the constant of 
proportionality is set by requiring that $\langle\rho\rangle=1$.  

However, if one views the transformation 
 $\rho \propto \exp(\delta_{\rm L})$ 
as a local deterministic mapping between the initial overdensity 
$\delta_{\rm L}$ and evolved density $\rho$ (see Coles \& Jones 1991 
and Smith, Scoccimarro \& Sheth 2007 for motivation), then 
equation~(\ref{eqn:pln}) is {\em not} the form which one obtains from 
the perturbation theory-like argument we have just described.  
In this context, equation~(\ref{eqn:pln}) corresponds to the 
Lagrangian rather than Eulerian space pdf.  
This curious fact does not appear to have received attention before.  
The distinction is important, because, for most power spectra of 
current interest, $\delta_{\rm L}(M|V)/\sigma_{\rm L}(M) \propto 
 \ln(M/V)/\sigma_{\rm L}(M)$ is not monotonic in $M$, so the method 
of Section~\ref{subsection:PT} breaks.  

\subsection{Example: Spherical collapse dynamics}\label{subsection:SC}
The spherical collapse model relates $\delta_{\rm L}$ and $\rho$; 
it is well approximated by  
\begin{equation}
 1+\delta_{\rm NL} \equiv \rho
  = \left(1-\frac{\delta_{\rm L}}{\delta_{\rm c}}\right)^{-\delta_{\rm c}}
 \label{eqn:SCapprox}
\end{equation}
(Bernardeau 1994). 
The exact value of $\delta_{\rm c}$ depends on the background 
cosmology.  It is $\delta_{\rm c}\approx 1.686$ in an 
Einstein de-Sitter model, and $\delta_{\rm c}\approx 1.66$ 
for the simulation results we present later.  As a result, 
setting $\delta_{\rm c} = 5/3$ is an excellent approximation 
which we will use to provide simple analytic examples of our 
results.  
Note that the model studied by Sheth (1998) was, in effect, 
$\delta_{\rm c}=1$; for white-noise initial conditions 
(initial spectral slope $n=0$), the resulting excursion set 
prediction is in exact agreement with HEPT.  As we discuss in 
more detail later, this expression with $\delta_{\rm sc}=3$ is 
what the Zeldovich approximation yields for the collapse of a 
spherical perturbation.   

The perturbation theory prediction for the pdf associated with 
the spherical collapse mapping (equation~\ref{eqn:SCapprox}) is 
\begin{eqnarray}
 \rho^2 p(\rho|V) &=& \frac{1}{\sqrt{2\pi \sigma_L^2(\rho)}}
  \exp \left[-\frac{\delta_L^2(\rho)}{2\sigma_L^2(\rho)} \right] \nonumber \\
 & &  \times \left(1-\frac{\delta_L(\rho)}{\delta_c}+\frac{\gamma}{6}\delta_L(\rho)\right). \label{eqn:unnormalizedPT}
\end{eqnarray}
The corresponding excursion set expression is 
\begin{eqnarray}
\rho^2 p(\rho|V) &=& \frac{1}{\sqrt{2\pi\sigma^2_L(\rho)}}\exp\left[-\frac{\delta^2_L(\rho)}{2\sigma^2_L(\rho)}\right] \nonumber \\
& & \times \left(1-\frac{\delta_L(\rho)}{\delta_c}+\frac{\gamma}{3}\delta_L(\rho) \right). \label{eqn:unnormalizedRW}
\end{eqnarray}
Note that this differs from the perturbation theory expression only 
because it has a factor of $\gamma/3$ rather than $\gamma/6$. 
For $|\delta_L(\rho)|\ll 1$, or for 
$\gamma\ll 6/\delta_{\rm sc} \approx 18/5$, 
this difference is negligible, and the two models will give very 
similar results.  Demonstrating this agreement between the two methods 
is one of the central results of this paper.  

\begin{figure}
 \begin{center}
 \includegraphics[scale=0.433]{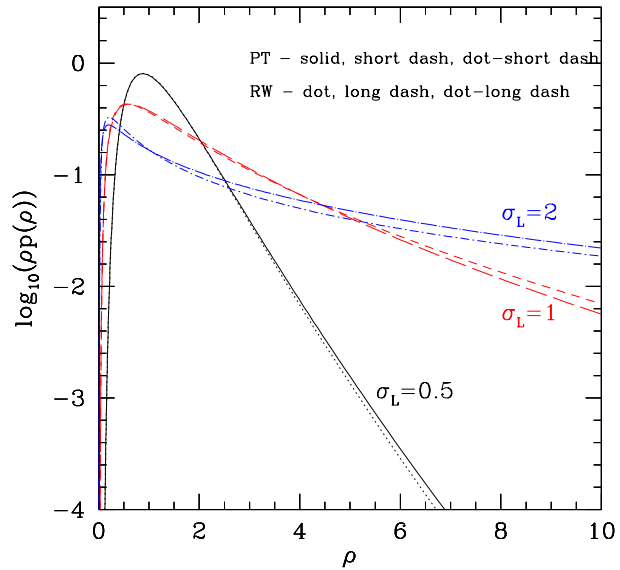}
 \caption{Comparison of equation~(\ref{eqn:besselPT}) (solid, 
          short-dashed, dot-short dashed) and 
          equation~(\ref{eqn:besselRW}) (dotted, long-dashed, dot-long 
          dashed) for $\sigma_{V}=0.5,1.0,2.0$ respectively.}
 \label{fig:besselk}
 \end{center}
\end{figure}

To illustrate the similarity, and to get a feel for the magnitude 
of the differences, Figure~\ref{fig:besselk} shows the nonlinear pdf 
predicted by these models when $\delta_c=5/3$ and the initial power 
spectrum was $P(k)\propto k^{-6/5}$.
The three sets of curves are for $\sigma_{\rm V} = 1/2, 1$,and $2$ 
(narrowest to broadest distributions).

\begin{figure*}
\begin{center}
\includegraphics[scale=0.8]{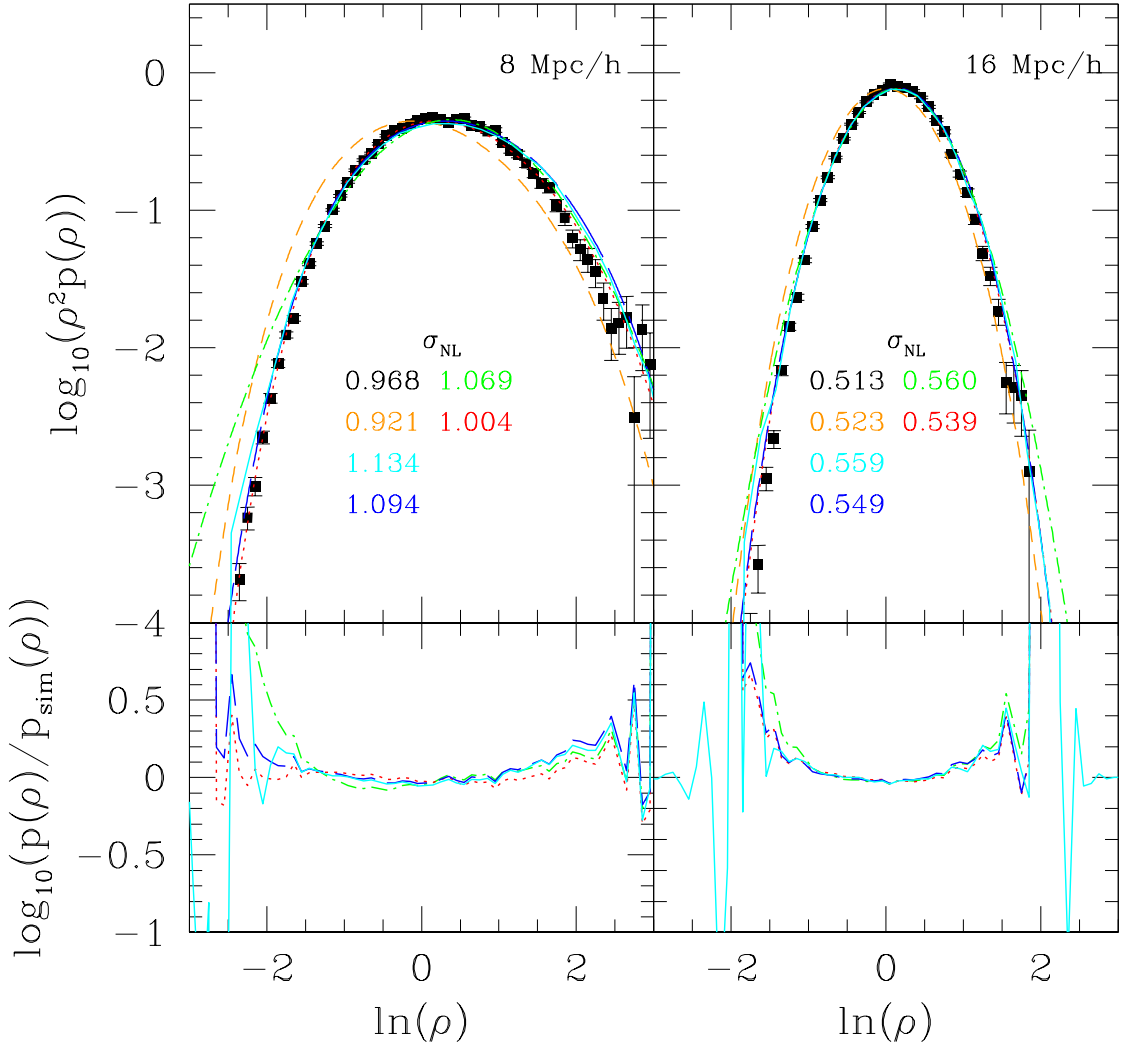}
\vspace{-0.6cm} 
\caption{Comparison of the measured $\rho^2 \,p(\rho)$ (filled squares) 
         with various spherical collapse based predictions for this 
         quantity. Top: dot-dashed (green) curves show the standard Lognormal 
         (equation~\ref{eqn:pln}).
         Short-dashed (orange) and dotted (red) curves show perturbation 
         theory 
         before (equation~\ref{eqn:unnormalizedPT}) and after the 
         normalization discussed in Section~\ref{norm} 
         (the dotted curve is simply shifted slightly to the right);
         solid (cyan) curves show the exact Monte-Carlo solution of the 
         excursion set model, and long-dashed (blue) curves show the 
         associated 
         analytic approximation (equation~\ref{eqn:unnormalizedRW}). 
         The nonlinear rms fluctuations of different models are shown 
         in the following order: the VLS simulation (0.968 and 0.513), 
         the Lognormal (1.069 and 0.560), 
         the perturbation theory before normalization (0.921 and 0.523), 
         the perturbation theory after normalization (1.004 and 0.539), 
         the exact Monte-Carlo soultion of the excursion set model (1.134 and 
         0.559), and the analytic approximation (1.094 and 0.549);
         Bottom: the normalized distributions and the Lognormal
         divided by the measurements.  }
\label{fig:lgrho2Prho}
\end{center}
\end{figure*}

\subsection{Comparison with simulations}\label{section:simulation}
We compare our predictions with measurements made in the Very Large 
Simulation (VLS) \cite{vls}.  The simulation box represents a cube 
$479h^{-1}\ {\rm Mpc}$ on a side in a cosmological model where 
($\Omega_m, \Omega_{\Lambda},h,\sigma_8) = (0.3, 0.7, 0.7, 0.9)$.
It contains $512^3$ particles, so the associated particle mass is 
$6.86\times 10^{10}h^{-1}\ {\rm M_\odot}$.  

We initially counted particles in cubic cells of side 
$(479/37)~h^{-1}$Mpc; 
the volume of each cell is equivalent to that of a sphere of radius 
$8.03~h^{-1}{\rm Mpc}$.  In the results which follow, counts in 
larger cells were obtained by summing up counts in neighbouring cells.  
For cells near the boundary of the box, we used the fact that the 
simulation was run using periodic boundary conditions.  

Because one does not simulate the nonlinear evolution of a 
continuous density field, rather, one simulates the motion of 
particles, the relation between the discrete particle pdf one 
measures in simulations is non-trivial.  Typically, one assumes 
that the distribution of particle counts in cells is 
\begin{equation}
 p(N|V) = \int dM \, p(M|V)\, p(N|M),
\end{equation}
where $p(N|M)$ denotes the probability that a mass $M$ is 
represented by $N$ particles.  If $m$ denotes the particle mass, 
then the Poisson model assumes that $p(N|M) = (M/m)^N\exp(-M/m)/N!$.  
Our comparisons with the theoretical models of the previous section 
assume that discreteness effects are negligible, so 
$N/\bar{N} =  M/\bar{M}$.

\begin{figure*}
\begin{center}
\includegraphics[scale=0.8]{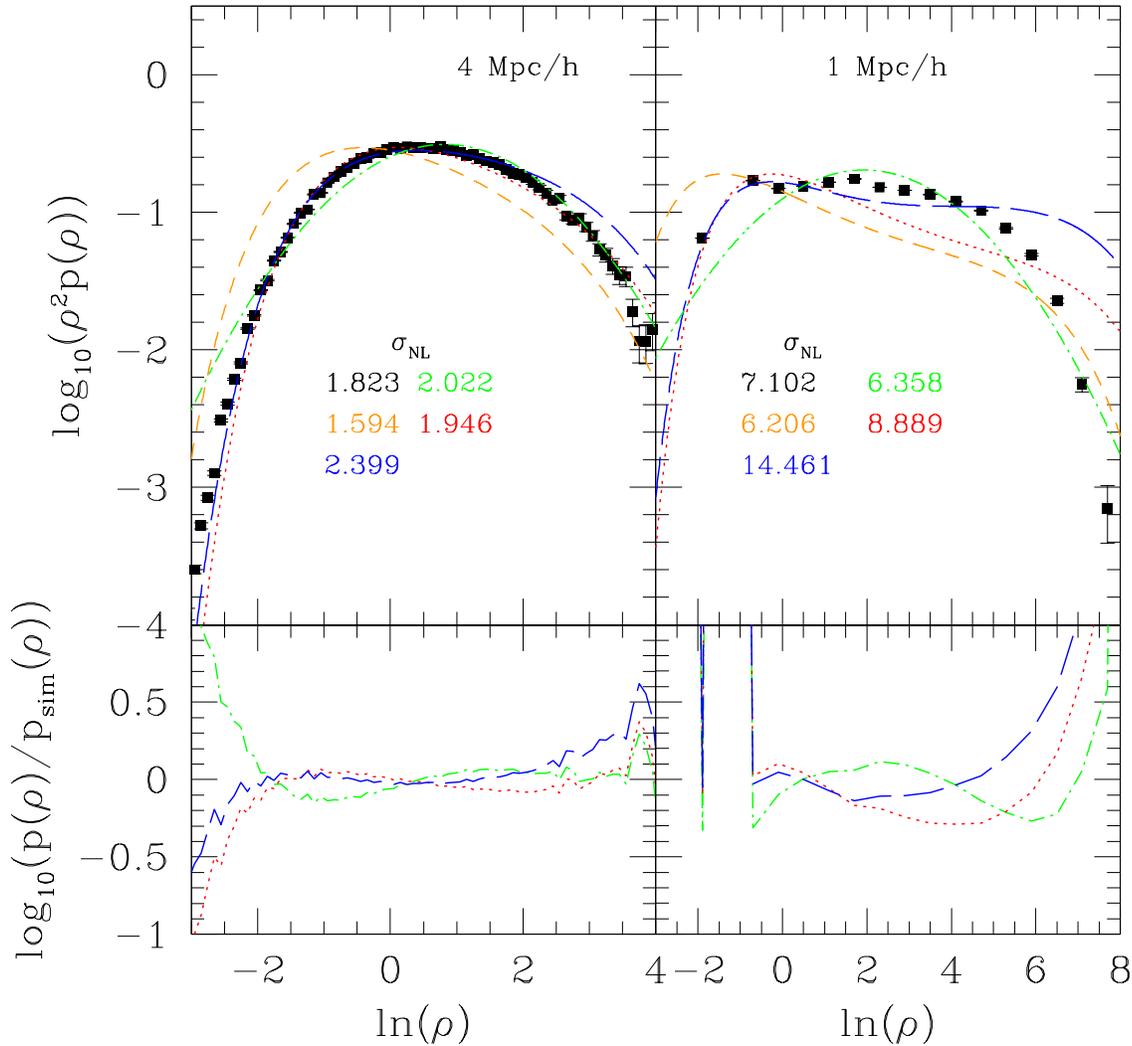}
\vspace{-0.6cm}
\caption{Same as previous figure, but for smaller cells, in which 
         the rms fluctuation is significantly larger than unity. The results 
         from the exact Monte-Carlo solution of the excusrion set model are 
         not included as the associated analytical approximation gives 
         very similar results.}
\label{fig:lgrho2Prho1-4}
\end{center}
\end{figure*}

The top panel of Figure~\ref{fig:lgrho2Prho} shows the pdf of the 
dark matter in cells of two different volumes (equivalent to spheres 
of radii $8$ and $16$ $h^{-1}$Mpc), chosen to have rms fluctuation 
values of unity or less.  
Black squares show the measurements in the VLS, where
 $\rho \equiv 1+\delta_{NL}= N/\bar{N}$.  
Curves show the predictions of the different models:  
the Lognormal is dot-dashed (this model has one free parameter, 
the nonlinear variance, which we compute from the fitting formula 
in Smith et al. (2003), rather than using the value measured in the 
simulations); 
raw and normalized perturbation theory model are short-dashed and 
dotted (note that the normalised curve is simply shifted to the 
right of the unnormalised curves); and 
the excursion set model is solid (long-dashed curve shows the 
analytic approximation). 
The nonlinear variances $\sigma$ associated with these various 
models are also shown (from left to right, top to bottom: 
VLS simulation, lognormal, unnormalised perturbation theory, 
normalised perturbation theory, excursion set from Monte-Carlo, 
and excursion set from the approximation formula).

To reduce the dynamic range, the bottom panel shows the predictions 
and Lognormal all ratioed to the measurements. 
This shows clearly that both the normalised perturbation theory  
and excursion set models are in better agreement with the simulations 
than is the Lognormal.  
We note in passing that Betancort-Rijo \& Lopez-Corredoira (2002) 
provide a relatively simple analytic expression for the nonlinear pdf.  
Their expression is, essentially, yet another prediction for the 
spherical evolution based pdf.  It fares substantially worse than 
either of our models - it overpredicts the high density tail - so we 
do not show it here.  

Figure~\ref{fig:lgrho2Prho1-4} shows that these spherical collapse 
based predictions remain accurate even on scales where the rms 
fluctuation is significantly larger than unity.  This is well beyond 
the regime where standard perturbation theory is expected to be 
valid, and indeed, the perturbation and excursion set predictions 
for the high density tails differ significantly.  Perturbation theory 
provides a substantially better description of the simulations when 
the rms fluctuation is of order 2 (left panel) but is worse when the 
rms is larger (right panel).  
On the other hand, the average density within the virial radius of 
a dark matter halo is about 200 times the background density; thus, 
in the regime where $\ln\rho\ge 5$, the mapping of 
equation~(\ref{eqn:SCapprox}) is suspect.  

Figures~\ref{fig:lgrho2Prho} and~\ref{fig:lgrho2Prho1-4} show 
that the spherical model produces rather accurate predictions 
for the shape of the nonlinear pdf, at least on scales where the 
nonlinear rms fluctuation is smaller than about 2.  We show later 
that this can be used to motivate an algorithm which uses the 
nonlinear density field to provide an estimate of the initial pdf.

\section{Stochastic mappings from initial to final density}\label{stochastic}
In this section we replace the assumption of a spherical collapse 
in favor of the ellipsoidal collapse model.  This evolution is 
considerably more complex, and so we only present results to first 
and second order in the dynamics.  We use the formulation of this 
model in which it reduces, to lowest order, to the Zeldovich 
Approximation \cite{bm96}.

\subsection{The Zeldovich Approximation}\label{zel}
In the Zeldovich Approximation, the nonlinear density is a deterministic 
function of three numbers in the initial distribution.  These are 
the eigenvalues $\lambda_i$ of the deformation tensor, a $3\times 3$ 
symmetric matrix whose elements are the second derivatives of the 
initial potential field.

The initial density $\delta_{\rm L}$ is the sum of the three 
eigenvalues, whereas the nonlinear density is 
\begin{equation}
 \rho = \prod_{i=1}^3 (1 - \lambda_i)^{-1}.
 \label{eqn:rhoZA}
\end{equation}
(note that $\rho \to 1 + \sum_i \lambda_i$ when $\lambda\ll 1$). 
All three eigenvalues are the same for a sphere, in which case 
$\lambda = \delta_{\rm L}/3$.  In this case, the relation above 
reduces to that of the spherical model from the previous section 
with $\delta_{\rm c}=3$.  Similarly, $\delta_{\rm c}=2$ is 
associated with regions where the smallest eigenvalue is zero, 
and the other two are each equal to $\delta_{\rm L}/2$.  If one 
calls such an object a filament, then a sheet has two eigenvalues 
equal to zero and the third equal to $\delta_{\rm L}$, so the 
evolution is given by equation~(\ref{eqn:SCapprox}) with 
$\delta_{\rm c}=1$.  

The evolved pdf of this model has 
\begin{equation}
 \rho\,p(\rho|V)\,{\rm d}\rho 
 = \int {\rm d}{\bm \lambda}\, p({\bm \lambda}|\sigma) \,
   \delta_{\rm D}\left(\rho = \prod_{i=1}^3 (1 - \lambda_i)^{-1}\right)
\end{equation}
\cite{pk93,hks00,brlc02}.  

For Gaussian initial conditions, the joint distribution of 
$p(\lambda_1,\lambda_2,\lambda_3|\sigma)$ is known \cite{grf}, 
and it is straightforward to evaluate the integral above by 
Monte Carlo methods.  In practice, this distribution is a function 
of $\lambda_i/\sigma$, so it is useful to think of the expression 
above as 
\begin{equation}
 \rho\,p(\rho|V)\,{\rm d}\rho 
 = \int {\rm d}{\bm \lambda}\, p({\bm \lambda}|1) \,
   \delta_{\rm D}\left(\rho = \prod_{i=1}^3 (1 - \sigma\lambda_i)^{-1}\right).
   \label{eqn:ZAeqn}
\end{equation}  
Notice that if $\sigma=\sigma_{\rm L}(\bar\rho V)\equiv\sigma_{\rm V}$, 
then 
\begin{equation}
 \int p(\rho|V)\,{\rm d}\rho 
 = \int {\rm d}{\bm \lambda}\, p({\bm \lambda}|1) \,
    \prod_{i=1}^3 (1 - \sigma_{\rm V}\lambda_i) = 1.
\end{equation}
This choice is made by Hui et al.~(2000); in this case, the 
normalization problems of Section~\ref{norm} do not arise.
However, the analysis of the previous sections suggests that this 
ignores the effects of smoothing, and that setting 
$\sigma=\sigma_{\rm L}(M)$ is almost certainly a better choice 
\cite{br91,pk93,brlc02}.

Notice that if any one of the eigenvalues equals unity then the 
density diverges, and that the density becomes negative if one 
and only one or all three eigenvalues exceed unity.
This happens often when $\sigma\ge 1$ (with probability 11\% when 
$\sigma=1$), thus restricting the use of this approximation to 
large scales where $\sigma$ is small.  
Accounting for the effects of smoothing will mitigate this somewhat, 
since $\sigma_{\rm M}\ll \sigma_{\rm V}$ when $\rho\gg 1$, where we 
have defined $\sigma_{\rm M}\equiv\sigma_{\rm L}(M)$ and 
$\sigma_{\rm V}\equiv \sigma_{\rm L}(\bar\rho V)$.  
For simplicity, on the few occasions when the density does go 
negative, we take the absolute value of the {\em rhs} of 
equation~(\ref{eqn:rhoZA}); there is no physical motivation for this 
choice, but, in practice, this happens sufficiently rarely that it 
does not affect our results. 

\begin{figure*}
\begin{center}
\includegraphics[scale=0.8]{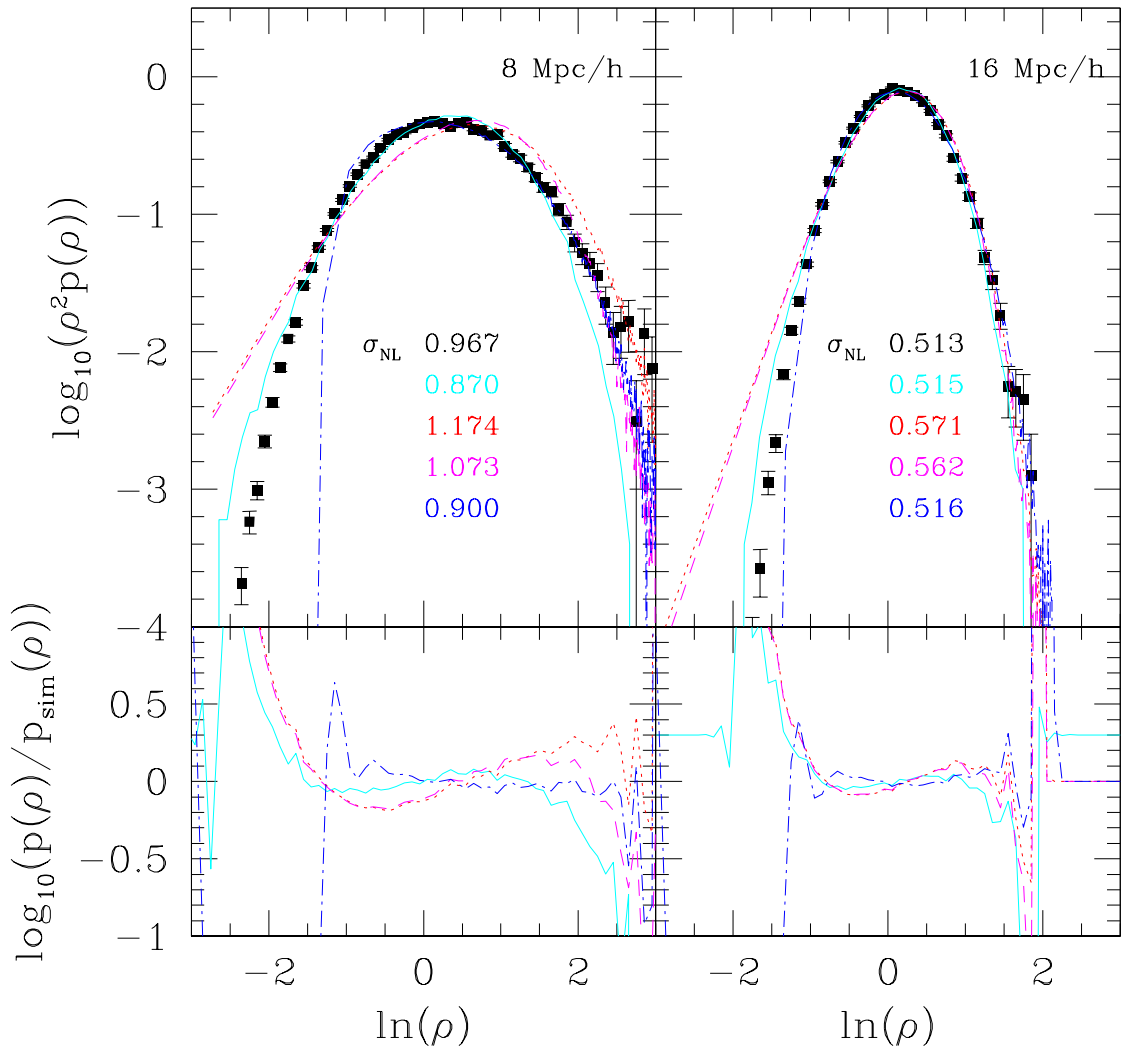}
\vspace{-0.6cm}
\caption{Comparison of the measured $\rho^2 \,p(\rho)$ (filled squares) 
         and various Zeldovich approximation-based predictions and second 
         order ellipsoidal model using perturbation-theory for this 
         quantity.  In the top panel red (dotted) and cyan (solid) curves 
         show the normalized 
         perturbation-theory and excursion-set predictions using the spherical 
         linear variance approximation. Magenta (dash) shows the 
         perturbation-theory prediction using the ellipsoidal variance 
         approximation (equation~\ref{eqn:ellvar}). Blue (dotted-dash) shows 
         the normalized perturbation-theory predictions using second order 
         ellipsoidal collapse model (equation~\ref{eqn:pdfEC}). 
         The numbers shown are the rms 
         values for (from top to bottom) VLS simulation, excursion-set 
         (spherical variance), perturbation-theory (spherical 
         variance), perturbation-theory (Zeldovich approximation using
         ellipsoidal variance), and perturbation-theory (second ellipsoid 
         collapse using ellipsoidal variance). Bottom panel shows the 
         corresponding distributions divided by the measurements.}
\label{fig:lgrho2PrhoZA}
\end{center}
\end{figure*}

The expression above is still not entirely self-consistent.  
Although it tries to account for the fact that the final size of a 
region is different from its initial size, it does {\em not} account 
for the fact that the shape has also changed \cite{br91,pk93,brlc02}. 
 If the final volume $V$ 
is spherical, the initial volume was not; the 
analysis above does not account for this.  However, this can be 
done in a relatively straightforward way by simply setting 
$\sigma_{\rm L}({\bm\lambda})$, where, for Eulerian spheres $V$ of 
radius $R_{\rm E}$ we require that $R_{\rm E} = R_i(1-\lambda_i)$.  
This means that the delta function picks out ellipsoids in the 
initial field which contained mass $M \propto \prod_{i=1}^3 R_i$, 
and which are now spheres of volume $V$ which contain the same mass 
$M$.  The function $\sigma_{\rm L}({\bm\lambda})$ is the rms 
fluctuation in the initial field when smoothed with a triaxial filter 
of shape given by $(R_1,R_2,R_3)$;  we have checked that it is well 
approximated by 
\begin{equation}
 \sigma^{\rm Ell}_{\rm L}({\bm \lambda}) = \sigma^{\rm Sph}_{\rm L}(M)\,
 \exp\left\{-\frac{B}{2} 
  \sum_{i<j}\left[\ln\left(\frac{1-\lambda_i}{1-\lambda_j}\right)\right]^2
 \right\},
 \label{eqn:ellvar}
\end{equation}
where $B=0.0486$, and the subscripts $ell$ and $sph$ stand for 
ellipsoidal and spherical respectively \cite{brlc02}.
This, then, is our perturbation theory based estimate associated with 
the Zeldovich Approximation.

Figure~\ref{fig:lgrho2PrhoZA} shows how the shape depedence of the 
rms fluctuation changes the predicted dark matter pdf. 
In the top panel the dotted curves are the predictions of 
perturbation theory using equation~(\ref{eqn:ZAeqn}) with 
$\sigma_{\rm L}(M)$ whereas the dashed curves use 
$\sigma_{\rm L}({\bm \lambda})$. The difference between these 
two is significant, meaning the shape dependence is important, 
in the high density tail ($\ln(\rho)>1$) of the $8h^{-1}$Mpc 
cells. The bottom panel compares the ratios of different models 
to the measurements. In neither cases does the Zeldovich Approximation 
agree well with the simulations: the difference is more than a factor 
of two in low density regions 
($\ln(\rho) < -2$ in $8h^{-1}$Mpc and $\ln(\rho) < -1.5$ in $16h^{-1}$ Mpc).

It is rather more complicated to account approximately for the 
effect of the change in shape in the excursion set approach.  
Our Monte Carlo algorithm works as follows.  
We generate $(\lambda_1,\lambda_2,\lambda_3)$ 
in each step of the walk following the procedure described by 
Sheth \& Tormen (2002).  
The variance associated with step $n$ is $S^{(n)}$.  This has an 
associated scale $R^{(n)}$.  We are interested in patches which today 
are spheres of radius $R_{\rm E}$.  (The extension to ellipsoids at 
the present time is straightforward).  These are those initial patches 
which satisfy $R_i^{(n)}(1-\lambda_i^{(n)}) = R_{\rm E}$, where 
$\lambda_i^{(n)}$ is the value of $\lambda_i$ after $n$ steps.  
We then want the largest mass associated with the various $R_i^{(n)}$.  
Since mass is proportional to $\prod_i R_i$, we want the largest 
value of $R_i^{(n)}$ for each $i$; in effect, we want the first 
crossing values $n_i$ for each of the three barriers 
$R_i^{(n)}(1-\lambda_i^{(n)}) = R_{\rm E}$.  Let $f(n_1,n_2,n_3)$ 
denote the fraction of walks which first cross the three barriers 
after $(n_1,n_2,n_3)$ steps.  Then the excursion set model sets 
\begin{equation}
 \rho\,p(\rho|V)\,{\rm d}\rho = f(M_{123})\, {\rm d}M_{123}.
\end{equation}
where 
\begin{equation}
 M_{123} \propto R_1^{(n_1)}R_2^{(n_2)}R_3^{(n_3)}
\end{equation}
denotes the mass associated with the first crossing at $(n_1,n_2,n_3)$.  
We then renormalize the left hand size following the discussion in 
Section~\ref{norm}. In Figure~\ref{fig:lgrho2PrhoZA} the cyan (solid) curves 
show the predictions of the excursion set approach. While they fit better 
than the perturbation theory's results the agreement is not very good at the 
low density region.

Although this algorithm allows for different shapes in the initial 
distribution, it only approximates the exact problem we wish to 
solve, because the $n$th step in any given direction is associated 
with the same value of $S_n$, whatever the values of $n$ in the 
other directions.  In effect, we are approximating $S(n_1,n_2,n_3)$ 
with the spherical value $S_{n_i}$ for each axis $i$.  

\subsection{The ellipsoidal collapse model}\label{EC}
The late time evolution in the ellipsodial collapse model is more 
complicated than in the Zeldovich Approximation \cite{bm96}.  
Nevertheless, we can still write the nonlinear density as a 
deterministic (albeit complicated) mapping of the three eigenvalues 
${\bm \lambda}$ in the initial field.  

Thus, the non-linear density is given by an expression that is 
analogous to equation~(\ref{eqn:rhoZA}).  Namely, 
\begin{equation}
\rho = \prod_{i=1}^{3} (1-\xi_i)^{-1},  \label{eqn:ecollapse}
\end{equation} 
where in an Einstein-de Sitter universe the evolution of $\xi_i$ is 
described by 
\begin{equation}
 a^2\frac{{\rm d}^2\xi_i}{{\rm d}a^2} + \frac{3}{2}a \,
 \frac{{\rm d}\xi_i}{{\rm d}a} = 
 \frac{3}{2}(1-\xi_i)\left(\frac{1}{3}\delta + \,
 \frac{1}{2} b_i\delta + L_{{\rm ext},i} \right),
\label{eqn:xievolution}
\end{equation}
with 
\begin{eqnarray}
 b_i & = & \frac{4}{15}(3\xi_i - \sum_j \xi_j) + O(\xi^2), \qquad {\rm and}\\
 L_{{\rm ext},i} & = & \lambda_i - \frac{1}{3}\sum_j \lambda_j
\end{eqnarray}
This expression for $L_{\rm ext}$ is known as the linear tide 
approximation \cite{bm96}.  
The initial condition of equation~(\ref{eqn:xievolution}) is the 
Zeldovich Approximation. 

The solution of the differential equation above can be written as a series:
 $\xi_i = \sum_j \zeta_i^{(j)}a^j$.  
In what follows we only consider the first two terms in this series:
\begin{eqnarray}
 \zeta_i^{(1)} & = & \lambda_i \\
 \zeta_i^{(2)} & = & \frac{3}{50}(I_1^2 -2I_2) + \frac{11}{175}I_2 
                      + \frac{3}{50}\lambda_i(2I_1 -5\lambda_i),
\end{eqnarray}
where $I_1 \equiv \sum_j \lambda_j$ and 
      $I_2 \equiv \sum_{j \neq k}\lambda_j\lambda_k$ 
(Ohta et al. 2004 provide a similar expansion).

We then apply the same argument as in the previous section for the 
perturbation-theory approach to calculate the pdf of the evolved 
density field.  Namely, we set 
\begin{equation}
\rho\, p(\rho|V)\,{\rm d}\rho = \int {\rm d}{\bm \lambda}\,p({\bm \lambda}|1)\,
    \delta_{\rm D}\left(\rho = \prod_{i=1}^3 (1 - \sigma\xi_i)^{-1}\right),
\label{eqn:pdfEC}
\end{equation}
where $\sigma\xi_i \equiv \sigma\zeta_i^{(1)} + \sigma^2\zeta_i^{(2)}$ and 
$\sigma$ includes the shape effect (equation~\ref{eqn:ellvar}, 
with $\xi_i$ values replacing $\lambda_i$ values). 

\begin{figure} 
\begin{center}
\includegraphics[scale=0.4]{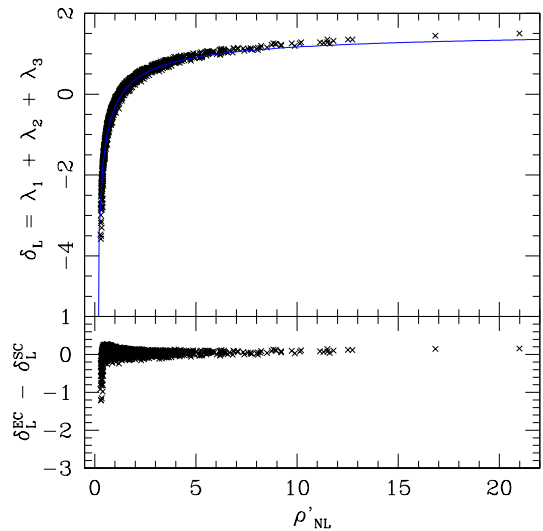}
\caption{Stochasticity in the mapping between linear and nonlinear 
         overdensity on scales of 8$h^{-1}$ Mpc. In the upper panel, 
         points show the ellipsoidal collapse mapping associated with 
         equation~(\ref{eqn:pdfEC}); solid line shows the 
         spherical collapse mapping of equation~(\ref{eqn:SCapprox}). 
         Both are normalised as described in Section~\ref{norm};
         $N_{\rm EC} = 1.087$ for ellipsoidal
         collapse and $N_{\rm SC} = 1.33$ for the spherical model.
         The lower panel shows the difference between these two mappings.}
\label{fig:ECmap}
\end{center}
\end{figure}

Figure~\ref{fig:ECmap} shows the correlation between the initial 
and final densities in this model for 8$h^{-1}$ Mpc cells.  
Recall that, because the evolution depends on all three $\lambda_i$, 
whereas the initial density $\delta_{\rm L}$ is simply proportional 
to $\sum_i\lambda_i$, we expect there to be some stochasticity in 
$\rho$ at each $\delta_{\rm L}$.  The Figure shows that this is not 
a very large effect.  Moreover, the solid line shows that the spherical 
collapse mapping (equation~\ref{eqn:SCapprox}) provides a reasonably 
good description of the mean mapping.  We exploit this fact in the 
next section.  

The dot-dashed lines in Figure~\ref{fig:lgrho2PrhoZA} show the 
predicted pdfs from this second order ellipsoidal collapse model.  
Note that inclusion of these second order terms provides a clear 
improvement over the Zeldovich Approximation prediction; this is 
true over all $\ln(\rho)> -1$ in both the 8 and 16$h^{-1}$ Mpc cells. 
Although the discrepancies at small $\rho$ are 
the most dramatic (we discuss these in the next section), we are 
actually more interested in the region around the peak of the pdf:  
the bottom panels show clearly that the Zeldovich Approximation 
curves lie snake around the simulation results, whereas this 
sideways S-shaped residual is largely absent in the second order 
ellipsoidal collapse model.

\section{Discussion}
We discussed two approaches (perturbation theory and excursion 
set) for calculating the dark matter pdf using two different 
models for the dynamics (spherical collapse and the Zeldovich 
Approximation) in each case.  Although both approaches are 
deterministic, in the sense that the nonlinear evolution is 
determined by the initial conditions locally, the excursion set 
is slightly less `local'.  We showed that both are expected to give 
similar predictions whenever the dynamics is approximated by local 
deterministic mappings (equation~\ref{eqn:PTpdf}), 
of which the spherical evolution model (equation~\ref{eqn:SCapprox}) 
is a special case (Figure~\ref{fig:besselk}).  The agreement is 
important, because the excursion set calculation allows one to 
connect discussions of the dark matter halo distribution with 
discussions of the pdf \cite{rks98}.  

Both the perturbation and excursion set approaches, when combined 
with the spherical evolution model, provide good descriptions of 
the nonlinear pdf seen in simulations (Figures~\ref{fig:lgrho2Prho}
and ~\ref{fig:lgrho2Prho1-4}).  
This agreement is slightly unexpected, in the sense that the 
spherical evolution model ignores the fact that the nonlinear 
evolution of a region may be determined by quantities other than 
its initial density.  The ellipsoidal collapse model is a specific 
example of this; we studied its first (Zeldovich Approximation) 
and second order expansions in some detail.  In this model, the 
evolution of a patch is determined by its overdensity as well as 
the surrounding shear field (only the overdensity matters for the 
spherical collapse model).  As a result, nonlinear evolution changes 
the shape as well as the volume of a patch (in the spherical 
collapse model, only the volume changes).  
We found that including the evolution of the shapes is important 
when the density is high, 
so this is particularly important when the cell size is small.  
This shape dependence also complicates the excursion set 
calculation (Section~\ref{zel}).  

Despite its increased complexity compared to the spherical evolution 
model, we found that the Zeldovich Approximation resulted in 
significantly worse agreement with the simulations 
(Figure~\ref{fig:lgrho2PrhoZA}).  Going to second order in the 
ellipsoidal collapse model resulted in more accurate predictions,
except in the lowest density regions. 
The relatively large discrepancies in underdense regions, 
in both models, may be related to the following fact.  A Taylor 
series in $\delta$ will converge more rapidly when $\delta>0$ than 
when $\delta<0$, since the latter will be an oscillating series.  
Thus, it may be that, to accurately represent the evolution of 
underdense regions, one must go beyond second order in the dynamics.  
Presumably, going to even higher order would further improve 
agreement with the simulations; this is the subject of work in progress.  

Although our perturbation theory calculation accounts for the 
evolution of shapes, our excursion set calculation does not.  
This too is the subject of on-going work.  
The excursion set calculation is particularly 
interesting in view of the fact that the marriage of ellipsoidal 
collapse with the excursion set approach appears to provide a 
substantially improved prediction of dark halo abundances \cite{smt01}.  
Finally, we note that our study of the real space pdf can be extended 
to redshift space; we are in the process of extending previous work 
on this problem \cite{ps97,hks00,ecpdf}.

\begin{figure*}
\begin{center}
\includegraphics[scale=0.75]{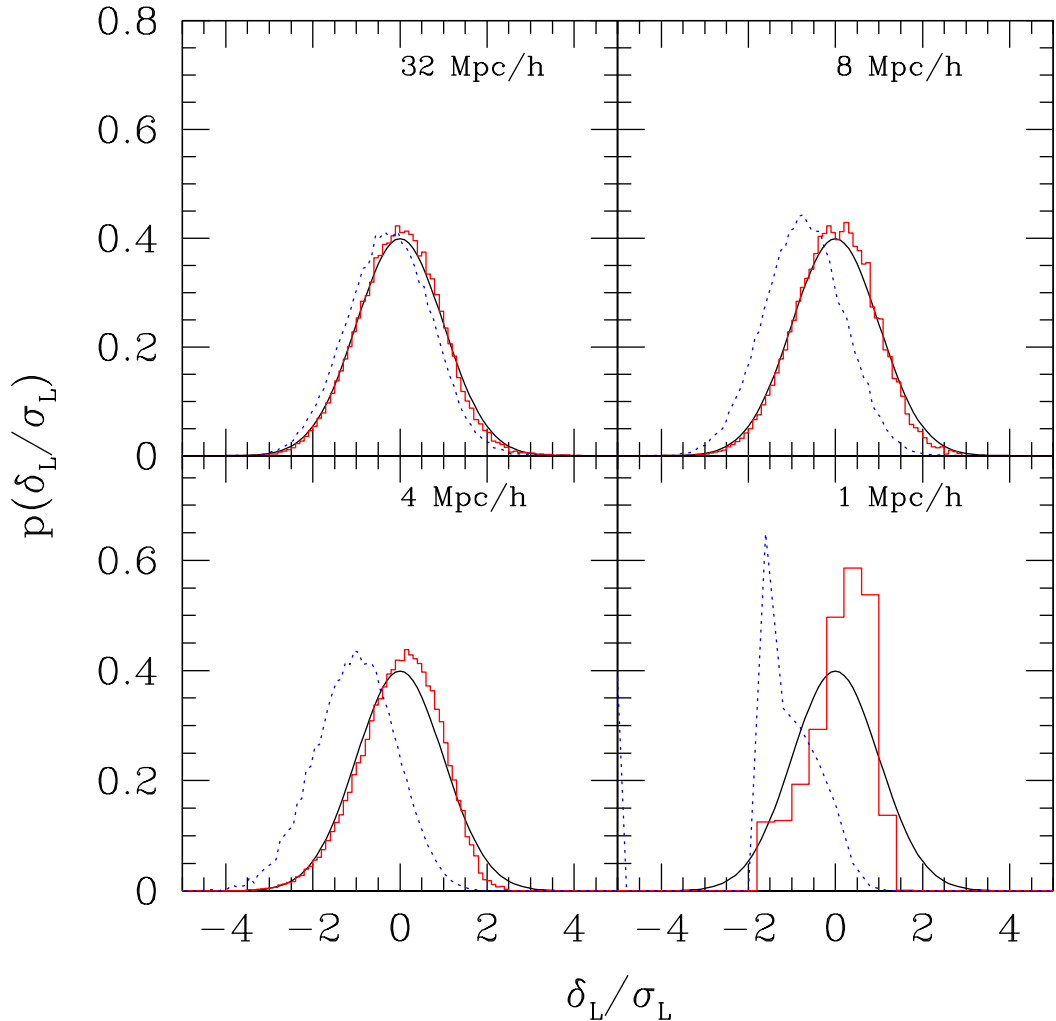}
\caption{Comparison of the reconstructed linear pdf of $\delta/\sigma$ 
         (histogram) with the expected zero-mean unit-rms Gaussian form 
         (solid line).  Different panels show results for reconstructions 
         from different smoothing scales.  The method works well for 
         large cells (top panels), but becomes increasingly inaccurate 
         for smaller cells (bottom panels).  When the nonlinear rms 
         fluctuation is larger than 2 (bottom right), then the 
         reconstructed pdf is rather different from the Gaussian form, 
         as one might expect from Figure~\ref{fig:lgrho2Prho1-4}.  
         Dotted curves show the bias which results from ignoring the 
         extra weighting factor associated with the transformation 
         from Eulerian to Lagrangian statistics.}
\label{fig:lineardm}
\end{center}
\end{figure*}

\subsection{An extension}
The accuracy of the spherical collapse based predictions, and 
the fact that the associated deterministic mapping from initial 
to final overdensity also provides a good description of the 
mean mapping in the ellipsoidal collapse model (Figure~\ref{fig:ECmap}), 
suggests that one might be able to estimate the shape of the initial 
pdf from a measurement of the evolved one as follows. Starting from 
equation~\ref{eqn:SCapprox} and the normalization method in section~\ref{norm},
 there is a mapping from the final density to its initial value.
For each measured $M = \bar\rho V(1+\delta_{\rm NL})$ set 
\begin{equation}
 \nu = \frac{1 - [(1+\delta_{\rm NL})/N_{\rm sc}]^{-1/\delta_{\rm c}}}
            {\sigma_{\rm L}(M/N_{\rm sc})/\delta_{\rm c}},
 \label{rescalenu}
\end{equation}
where $\delta_{\rm c}$ is the critical linear density associated 
with collapse in the spherical model, and $N_{\rm sc}$ is the 
corresponding normalization factor (Section~\ref{norm}).  
Then make a histogram of $\nu$, but rather than having each cell 
count equally, weight each by its value of 
$(1+\delta_{\rm NL})/N_{\rm sc}$.  
The resulting distribution of $\nu$ should provide a good estimate 
of the shape of the initial pdf of $(\delta_{\rm L}/\sigma_{\rm L})$.  

Figure~\ref{fig:lineardm} shows the results of this algorithm for 
a number of different scales.  Reconstruction of the initial pdf 
from the nonlinear pdf measured on scales larger than $8h^{-1}$Mpc 
works rather well (top panels).   The reconstructed distributions 
trace the same Gaussian shape very well; for reference, the smooth 
black curve shows a Gaussian with zero mean and unit variance. 
This mapping works well even when the cell size is as small as 
$4h^{-1}$Mpc (bottom left), but it fails at $1h^{-1}$Mpc (bottom right).
The Figure also shows that it is important to account for the factor 
of $1+\delta_{\rm NL}$ which relates Eulerian and Lagrangian statistics: 
weighting each cell in the nonlinear distribution equally would have 
lead one to conclude incorrectly that the reconstruction is biased 
even on very large scales.  Studying whether or not this is relevant 
for the MAK reconstruction method \cite{mak} is the subject of work 
in progress.  

In effect, the spherical evolution model, when combined with the 
nonlinear pdf, is able to provide a good description of the initial 
pdf for scales where the rms fluctuation is smaller than about 2. 
In principle, this might be used as the basis for a test of 
the Gaussianity of the initial conditions because the algorithm 
makes no explicit assumption about the form of the distribution 
of $\nu$.  The fact that the nonlinear pdfs from a range of different 
scales all map back to the same zero mean, unit variance Gaussian curve 
suggests that this method is a good test of Gaussianity. 
Although this is not the first method to reconstruct the shape of 
the initial one-point density field, it is simple and actually fares 
rather well compared to a number of previous methods 
(e.g. Kofman et al. 1994; Narayanan \& Croft 1999).  

This procedure is not as good a test of non-Gaussianity, in the 
sense that it does not yield a simple estimate of the shape of 
the initial distribution if it were non-Gaussian.  
This is because the reconstruction requires an estimate of the 
normalization factor $N_{\rm sc}$, and to calculate it, one must 
first specify the form of the initial distribution.  So the algorithm 
above is really a self-consistency check:  it checks if the assumed form 
of the initial pdf is consistent with that reconstructed from the 
measured nonlinear pdf, under the assumption that spherical evolution 
is a good model whatever the initial fluctuation field.  Of course, 
rescaling to $\nu$ involved a rescaling and transformation of the 
measured $M$, but it also involved $\sigma(M)$; this is a consequence 
of the fact that, for Gaussian fluctuations, information about the 
volume enters only through the scale dependence of the rms value.  
For more general distributions this may not remain true, in which 
case the mapping to the initial variable $\delta_{\rm L}$ may be 
more complicated than equation~(\ref{rescalenu}).  
Nevertheless, Figure~\ref{fig:lineardm} suggests that this may be 
an interesting avenue to explore in future work.  We are in the 
process of extending this method to include redshift space effects, 
as well as galaxy bias.  We believe it may be interesting to merge 
it with the method recently proposed by Eisenstein et al. (2007) for 
reconstructing the Baryon Acoustic Oscillation feature in the galaxy 
distribution.

\section*{Acknowledgements}
We thank R. Smith for help with the simulation data, 
R. Scoccimarro for discussions about stochasticity, 
E. Gazta\~naga for conversations during the NSF-PIRE School 
held in Santiago, Chile in March 2007, 
the organizers of the School for inviting us to participate, 
B. Jain for support, 
J. Betancort-Rijo for discussions, and 
the Aspen Center for Physics for its hospitality when some 
aspects of this work were completed.  
We would also like to thank the anonymous referee for a perceptive 
and helpful report.  
This work was supported in part by NSF grants 0507501 and 0520677.

\appendix
\section{Fully analytic pdf in the spherical model}
For a power-law power spectrum ($P(k)\propto k^n$), the spherical 
model allows a semi-analytical form of the pdf for some special 
choices of $n$.  
To see this clearly, define $\tilde{\delta}_L$ by 
\begin{equation}
\frac{\delta_L(\rho)}{\sigma_L(\rho)}=\frac{1}{\sigma_{V}}\delta_c\rho^{(n+3)/6}(1-\rho^{-1/\delta_c})\equiv \frac{\tilde{\delta}_L(\rho)}{\sigma_{V}},
\end{equation} 
where $\sigma_{V}$ is independent of $\rho$. If we set 
\begin{equation}
\frac{n+3}{6} = \frac{1}{2\delta_c},
\end{equation}
then 
\begin{eqnarray}
\rho & = & \left(\frac{\tilde{\delta}_L}{2\delta_c}+\sqrt{(\frac{\tilde{\delta}_L}{2\delta_c})^2+1}\right)^{2\delta_c} \\
N & = & \int^{\infty}_{-\infty}d\tilde{\delta}_L\frac{e^{-\tilde{\delta}_L^2/2\sigma^2_{V}}}{\sqrt{2\pi}\sigma_{V}}\left(\frac{\tilde{\delta}_L}{2\delta_c}+\sqrt{(\frac{\tilde{\delta}_L}{2\delta_c})^2+1}\right)^{-2\delta_c}.
\end{eqnarray}
and the normalised perturbation theory pdf is 
\begin{equation}
 \rho^2\,p(\rho) = \frac{\kappa(\rho/N,\delta_{\rm c})}
                           {2\sqrt{2\pi\sigma^2_{V}}}
 \exp\left[-\frac{\delta_c^2}{2\sigma^2_{V}}\,
                 \kappa^2(\rho/N,\delta_{\rm c})\right],
 \label{eqn:besselPT}
\end{equation}
where 
\begin{equation}
 \kappa(\rho,\delta_{\rm c}) = \rho^{1/2\delta_c} + \rho^{-1/2\delta_c}.
\end{equation}
Setting $\delta_c=5/3$ means $n=-6/5$ and so 
\begin{equation}
 N =\frac{5}{3}\frac{{\rm e}^{25/9\sigma^2_{V}}}{\sqrt{2\pi}\sigma_{V}}
 \left[K_{7/6}\left(\frac{25}{9\sigma^2_{V}}\right) + 
       K_{13/6}\left(\frac{25}{9\sigma^2_{V}}\right)\right],
\end{equation}
were $K_\nu(x)$ is a modified Bessel function of the second kind. 
The random walk prediction for this model is 
\begin{equation}
 \rho^2\,p(\rho) = \frac{(\rho/N)^{3/10}}{\sqrt{2\pi\sigma^2_{V}}}
  \exp\left[-\frac{25}{18\sigma^2_{V}}\,
             \kappa^2(\rho/N,\delta_{\rm c})\right]
 \label{eqn:besselRW}
\end{equation}
where
\begin{equation}
 N = \frac{10}{3}\,\frac{{\rm e}^{25/9\sigma^2_{V}}}{\sqrt{2\pi}\sigma_{V}}
     \,K_{7/6}\left(\frac{25}{9\sigma^2_{V}}\right).
\end{equation}
In this context, it is interesting that the family of models 
studied by Sheth (1998) included modified Bessel functions of 
the third, rather than second, kind.  
The solid and dotted curves in Figure~\ref{fig:besselk} show 
equations~(\ref{eqn:besselPT}) and~(\ref{eqn:besselRW}) for a 
few values of $\sigma_{V}$.

\label{lastpage}

\begin{thebibliography}{99}
\bibitem[Bernardeau 1994]{b94}
 Bernardeau F., 1994, A\&A, 291, 697
\bibitem[Bernardeau et al. 2002]{ptreview}
 Bernardeau F., Colombi S., Gazta\~naga E., Scoccimarro R., 2002, 
 Phys. Rep. 367, 1
\bibitem[Betancort-Rijo 1991]{br91}
 Betancort-Rijo J., 1991, MNRAS, 251, 399
\bibitem[Betancort-Rijo \& Lopez-Corredoira 2002]{brlc02}
 Betancort-Rijo J., Lopez-Corredoira M., 2002, ApJ, 566, 623
\bibitem[Bond \& Myers 1996]{bm96}
 Bond J.~R., Myers S.~T., 1996, ApJS, 103, 1 
\bibitem[Coles \& Jones 1991]{cj91}
 Coles P., Jones B. J. T., 1991, MNRAS, 248, 1
\bibitem[Doroshkevich 1970]{grf}
 Doroshkevich A. G., 1970, Astrofizika, 6, 581
\bibitem[Eisenstein et al. 2007]{}
 Eisenstein D. J., Seo H.-J., Sirko E., Spergel D. N., 2007, ApJ, 664, 675
\bibitem[Gazta\~naga \& Croft 1999]{gc99} 
 Gazta\~naga E., Croft R. A. C., 2001, MNRAS, 309, 885
\bibitem[Hui, Kofman \& Shandarin 2000]{hks00} 
 Hui L., Kofman L., Shandarin S. F., 2000, ApJ, 537, 12 
\bibitem[Kofman et al. 1994]{kbg94}
 Kofman L., Bertschinger E., Gelb J.~M., Nusser A., Dekel A., 1994, 
  ApJ, 420, 44
\bibitem[Makler et al. 2001]{mkc01}
 Makler M., Kodama T., Calv\~ao M. O., 2001, ApJ, 557, 88
\bibitem[Mohayaee et al. 2006]{mak}
 Mohayaee R., Mathis H., Colombi S., Silk J., 2006, MNRAS, 365, 939
\bibitem[Narayanan \& Croft 1999]{nc99}
Narayanan V.~K., Croft R.~A., 1999, ApJ, 515, 471
\bibitem[Ohta et al. 2004]{ecpdf}
 Ohta Y., Kayo I., Taruya A., 2004, ApJ, 608, 647
\bibitem[Padmanabhan \& Subramanian 1993]{pk93}
 Padmanabhan T., Subramanian K., 1993, ApJ, 410, 482
\bibitem[Protogeros \& Scherrer 1997]{ps97} 
 Protogeros Z. A. M., Scherrer R. J., 1997, MNRAS, 284, 425
\bibitem[Scherrer \& Gazta\~naga 2001]{sg01} 
 Scherrer R. J., Gazta\~naga E., 2001, MNRAS, 328, 257
\bibitem[Scoccimarro \& Frieman 1999]{hept} 
 Scoccimarro R., Frieman J., 1999, ApJ, 520, 35
\bibitem[Sheth 1998]{rks98} Sheth R. K., 1998, MNRAS, 300, 1057
\bibitem[Sheth \& Tormen 1999]{st99} 
 Sheth R. K., Tormen G., 1999, MNRAS, 308, 119
\bibitem[Sheth, Mo \& Tormen 2001]{smt01}
 Sheth R.~K., Mo H., J., Tormen G., 2001, MNRAS, 323, 1
\bibitem[Sheth \& Tormen 2002]{st02}
 Sheth R.~K., Tormen G., 2002, MNRAS, 329, 61 
\bibitem[Sheth \& van de Weygaert 2004]{voids} 
 Sheth R. K., van de Weygaert R., 2004, MNRAS, 350, 517
\bibitem[Smith et al. 2003]{smithetal}
 Smith R. E., Peacock J.~A., Jenkins A., White S.~D.~M., Frenk C.~S., 
 Pearce F.~R., Thomas P.~A.,Efstathiou G., Couchman H.~M.~P., 2003,
 MNRAS, 341, 1311
\bibitem[Smith et al. 2007]{sss07}
 Smith R. E., Scoccimarro R., Sheth R. K., 2007, PRD, accepted
\bibitem[White \& Silk 1979]{ws79}
 White S.~D.~M., Silk J., 1979, ApJ, 231, 1 
\bibitem[Yoshida et al. 2001]{vls} 
 Yoshida N., Sheth R. K., Diaferio A., 2001, MNRAS, 328, 669
\bibitem[Zeldovich 1970]{zel70}
 Zeldovich Y. B., 1970, A\&A, 5, 84
\end{thebibliography}
\end{document}